\pdfoutput=1
\documentclass{aa}

\usepackage{graphicx}
\usepackage[varg]{txfonts}
\usepackage{graphicx}
\usepackage{amsfonts,amsmath,amssymb}
\usepackage{url}
\usepackage{natbib}

\begin{document}

\title{TYC 8241 2652 1 and the case of the disappearing disk: no smoking gun yet
\thanks{Based on observations made with ESO telescopes at the Paranal Observatory 
(ESO program IDs 090.C-0697(A), 090.C-0904(A), and 095.C-0438(A)) and on observations obtained with XMM-Newton, an ESA science mission with instruments and contributions directly funded by ESA Member States and NASA.}}

\author{Hans Moritz G\"{u}nther\inst{\ref{MIT}} \and Stefan Kraus\inst{\ref{Exeter}} \and Carl Melis\inst{\ref{SanDiego}} \and Michel Cur\'e\inst{\ref{Valparaiso}} \and Tim Harries\inst{\ref{Exeter}} \and Michael Ireland\inst{\ref{ANU}} \and Samer Kanaan\inst{\ref{Valparaiso}} \and Katja Poppenhaeger\inst{\ref{Belfast},\ref{CfA}} \and Aaron Rizzuto\inst{\ref{Austin}} \and David Rodriguez\inst{\ref{Santiago}} \and Christian P.\ Schneider\inst{\ref{ESA}} \and Michael Sitko\inst{\ref{Bolder}} \and Gerd Weigelt\inst{\ref{MPfR}} \and Matthew Willson\inst{\ref{Exeter}} \and Scott Wolk\inst{\ref{CfA}}}
 
\institute{
Massachusetts Institute of Technology, Cambridge, MA, USA\email{hgunther@mit.edu}\label{MIT}
\and
University of Exeter, School of Physics, Astrophysics Group, Stocker Road, Exeter, EX4 4QL, UK\label{Exeter}
\and
Center for Astrophysics and Space Sciences, University of California, San Diego, California 92093-0424, USA\label{SanDiego}
\and
Instituto de F\'isica y Astronom\'ia, Universidad de Valpara\'iso, Valpara\'iso, Chile\label{Valparaiso}
\and
Research School of Astronomy and Astrophysics, Australian National University, Canberra, ACT 2611, Australia\label{ANU}
\and
Queen's University Belfast, Astrophysics Research Centre, Belfast, BT7 1NN, United Kingdom\label{Belfast}
\and
University of Texas at Austin, 2515 Speedway, Austin, TX 78712, USA\label{Austin}
\and
Departamento de Astronomia, Universidad de Chile, Casilla 36-D, Santiago, Chile\label{Santiago}
\and
ESTEC/ESA, Keplerlaan 1, 2201 AZ, Noordwijk, The Netherlands\label{ESA}
\and
Center for Extrasolar Planetary Studies, Space Science Institute, Boulder, CO 80301, USA\label{Bolder}
\and
Max Planck Institute for Radio Astronomy, Auf dem H\"ugel 69,  D-53121 Bonn, Germany\label{MPfR}
\and
Harvard-Smithsonian Center for Astrophysics, 60 Garden Street, Cambridge, MA 02138, USA\label{CfA}
}

\date{Received date / Accepted date }

\abstract{TYC\;8241~2652~1 is a young star that showed a strong mid-infrared (mid-IR,
  8-25 $\mu$m) excess in all observations before 2008 consistent with
  a dusty disk. Between 2008 and 2010 the mid-IR luminosity of this system
  dropped dramatically by at least a factor of 30 suggesting a loss of dust
  mass of an order of magnitude or more.}
 {We aim to constrain possible models including removal of disk material by
   stellar activity processes, the presence of a binary companion, or
     other explanations suggested in the literature.}
 {We present new X-ray observations, optical spectroscopy, near-IR
   interferometry, and mid-IR photometry of this system to constrain its
   parameters and further explore the cause of the dust mass loss.}
{In X-rays TYC\;8241~2652~1 has all properties expected from a young star: Its
  luminosity is in the saturation regime and the abundance pattern shows
  enhancement of O/Fe. The photospheric H$\alpha$ line is filled with
  a weak emission feature, indicating chromospheric activity consistent
  with the observed level of coronal emission. Interferometry does not detect a companion and sets upper limits on the companion mass of 0.2, 0.35, 0.1 and 0.05 $M_{\odot}$ at projected physical separations of 0.1-4~AU,4-5~AU, 5-10~AU, and 10-30~AU, respectively (assuming a distance of 120.9~pc). Our mid-IR measurements, the first of the system since 2012, are consistent with the depleted dust level seen after 2009.}
{The new data confirms that stellar activity is unlikely to destroy the
    dust in the disk and shows that scenarios where either TYC\;8241~2652~1
    heats the disk of a binary companion or a potential companion heats the
    disk of  TYC\;8241~2652~1 are unlikely.}


\titlerunning{TYC 8241 2652 1 and the case of the disappearing disk}
\authorrunning{H. M. G\"unther et al.} 

\maketitle

\section{Introduction}

Many young stars are surrounded by circumstellar disks. In the early phases of star formation, the gas from the parent molecular cloud collapses and the conservation of angular momentum leads to the formation of an accretion disk around the central object. These disks contain a significant amount of dust, but by far the largest fraction of their mass is contained in the disk gas, mostly hydrogen and helium. The upper layer of the dust intercepts some of the stellar emission and re-radiates it in the infra-red (IR) wavelength range. Such disks thus are often discovered through surveys in the mid-IR. Some of the mass is accreted onto the central star, causing a  veiling of the stellar photospheric spectrum, bright hydrogen emission lines and an excess of soft X-ray emission \citep[for a review see][]{2011AN....332..448G}.
Furthermore, mass is lost from the disk through disk winds or photoevaporation and dust in the disk can coagluate to form larger grains, pebbles and eventually planetesimals and planets, which may be massive enough to accrete gas from the disk, too. These mass loss processes lead to disk life times of a few Myrs \citep{2001ApJ...553L.153H,2007ApJ...671.1784H}.

After the initial disk is gone, some young stars develop a debris disk. Planets or planetesimals in the disk collide, releasing a stream of small dust grains, which once again re-radiate the stellar emission in the IR. Those disks are almost gas-free \citep{2006ApJ...651.1177P,2007ApJ...668.1174F,2010A&A...516A...8S} and the small dust grains will be blown out of the system through radiation pressure even for stars of modest luminosity; for the continued presence of a dust disk, the grains need to be replenished through continued collisions \citep[for a review see][]{2008ARA&A..46..339W}.

The time scale for planet formation and dispersal of primordial disks is Myrs
but the disk emission can change faster. In general, observations at
shorter wavelength probe the disk at smaller radii. In the near-IR we see the
relatively warm, innermost parts of a disk. With changes of the accretion
state, the inner radius can fluctuate on time scales of hours to days changing the amount of near-IR excess above the stellar photosphere. In the most extreme cases, the inner disk hole can grow so large that no near-IR excess remains. Indeed, \citet{2012ApJ...755...65R} observe transient IR excesses in one third of their sample in the $K$ band. At longer wavelengths, the disk emission changes following changes of the illumination, e.g. with the luminosity of the accretion spot (hours), or when the accretion spot on the star rotates in and out of view (days). Monitoring of several star forming regions at 3.6 and 4.5~$\mu$m by the YSOVAR \citep{Rebull_2014} and CSI 2264 \citep{2014AJ....147...82C} projects reveal a plethora of lightcurves with variability between hours and weeks. However, out of $\sim$11,000 objects monitored in the YSOVAR project over several years, no more than 2 have data that indicates that their mid-IR excess vanishes completely, which would be compatible with losing their disk within a few years \citep{Rebull_2014}. 

In debris disks the grains need to be replenished constantly \citep{2008ARA&A..46..339W}, but once there is a collisional cascade in place that grinds down larger objects into sub-mm dust, there is little reason for the dust production to cease suddenly. The IR luminosity of debris disks thus follows the evolution of the larger bodies which ultimately supply the mass; the typical life time of debris disks is 100s of Mys to Gyrs \citep[e.g.][]{2009ApJ...697.1578G,2009ApJS..181..197C,2013ApJ...768...25G}.

Only three objects with strong variability in the mid-IR are known, and \object{TYC 8241 2652 1} \citep{2012Natur.487...74M}, the focus of this paper, is the most extreme case; the other two are the two debris disk stars ID8, and HD~23514 \citep{2012ApJ...751L..17M}. TYC\;8241~2652~1 used to show a normal K2 stellar spectrum in the optical and near-IR (out to about 5 $\mu$m) and a strong excess over the photospheric spectrum in the mid-IR (8-25 $\mu$m) indicative of either a primordial or a debris disk. Between May 2008 and January 2010 the mid-IR luminosity of TYC\;8241~2652~1 dropped by at least a factor of 30, but no changes were observed at shorter wavelengths. The presence of a Lithium absorption line proves that the star is young \citep{2012Natur.487...74M}. Its distance of $120.9 \pm 5.4$~pc \citep{2016arXiv160904303L} and its space motion are consistent with membership in either the Lower-Centaurus-Crux association (10-20 Myr old) or the TW Hydrae association (TWA, ~8 Myr) \citep[see supplementary informaion of][and references therein for details]{2012Natur.487...74M}. The photospheric H$\alpha$ line is filled, but not in emission, indicating stellar activity, but no or only very weak accretion, which is consistent with little or no gas reservoir in the disk. 

Based on the age and the spectral energy distribution, \citet{2012Natur.487...74M} suggest that this is a debris disk. Alternatively, the object might have hosted a primordial gas-rich disk in an unusual and very short-lived late stage of disk dispersal, where only parts of the inner region of the disk remain to be depleted.
So far, no scenario that is consistent with all observational constraints has been able to explain the drop in IR flux. \citet{2012Natur.487...74M} discuss several possibilities that we summarize here (see the supplementary information of their article for more details and references): A change of the disk structure at the inner edge would cast shadows on the outer disk and reduce the mid-IR luminosity, but the energy will still be re-radiated, just at shorter wavelengths. No such excess is observed. Only a shadowing of the whole system by some intervening material could cause the whole disk to dim; however, the optical flux of the central star remains unchanged. In a second class of explanations, material is physically removed from the disk. Options are (i)~a collisional avalanche that grinds the larger dust grains down until they are efficiently removed by stellar radiation, (ii)~runaway accretion (a gas-driven instability that suddenly empties the disk onto the host star), and (iii)~massive X-ray flares that evaporate the dust grains. The collisional avalanche and the runaway accretion models $-$ while capable of producing comparable timescales to the limits for the disappearance event $-$ are dynamically very challenging to realize; the X-ray evaporation would require a flare that outshines the bolometric luminosity of its host star by several orders of magnitude. No such flare has ever been observed for solar-like stars \citep[see supplementary informaion of][and references therein for details on these models]{2012Natur.487...74M}. \citet{2013ApJ...765L..44O} point out that large flares are associated with coronal mass ejections, which are more efficient in removing dust grains than the X-ray photons themselves and thus smaller, fortuitously-aligned flares could be sufficient to reduce the dust mass in the disk.

Given the open questions about which scenario presented $-$ if any $-$ is appropriate for the TYC\,8241~2652~1 system, we obtained new observations that aim to better characterize its circumstellar environment and further test models for its dramatic drop in mid-IR flux. Because TYC\;8241~2652~1 is the most extreme case of IR variability known to date, it sets the strictest limits on the processes in the disk that are responsible for the removal of warm dust. 

In section~\ref{sect:obs} we present the new X-ray, optical and IR observations and the associated data reduction. Section~\ref{sect:results} shows the results we can extract from that data. We discuss implications in section~\ref{sect:discussion} and end with a short summary in section~\ref{sect:summary}.

\section{Observations}
\label{sect:obs}

In this section we describe observations analyzed in this article ordered by wavelength: X-ray (Sect.~\ref{sect:xrayobs}), optical spectroscopy (Sect.~\ref{sect:optobs}), near-IR interferometry (Sect.~\ref{sect:nearirobs}), and mid-IR imaging (Sect.~\ref{sect:midirobs}). They are summarized in table~\ref{tab:obs} with the wavelength or wavelength range of the most relevant features and the phenomenon we wanted to probe with these observations.

\begin{table*} 
\caption{{Observations\label{tab:obs}}}
    \begin{tabular}{lrlll}
        band & wavelength & Instrument & data analysis & purpose\\ 
        \hline\hline
        X-ray & 0.1-2.4~keV & ROSAT & photometry & stellar activity\\
        X-ray & 0.2-3~keV & XMM-Newton & photometry & stellar activity\\
        X-ray & 0.3-3~keV & Chandra & spectroscopy & stellar activity\\
        optical & 656~nm & SSO/WiFeS & spectroscopy & stellar activity / accretion\\
        optical & 656~nm & SOAR/Goodman & spectroscopy & stellar activity / accretion\\
        near-IR & 2.2~$\mu$m & VLT/AMBER & interferometry & search for companions\\ 
        near-IR & 3.8~$\mu$m & VLT/NACO & interferometry & search for companions\\
        mid-IR & 10.7~$\mu$m & VLT/VISIR & photometry & return of dust?\\
        mid-IR & 11.4~$\mu$m & Subaru/COMICS & photometry & return of dust?\\
        \hline
    \end{tabular} 
\end{table*}

\subsection{X-ray observations}
\label{sect:xrayobs}
We compare data on TYC\;8241~2652~1 from three different X-ray missions: A source at a distance of 20\arcsec{} from the optical position of TYC\;8241~2652~1 -roughly the positional uncertainty- is detected in the \emph{ROSAT} All Sky Survey (RASS) bright source catalog \citep{1999A&A...349..389V}. The source \object{1RXS J120900.4-512050} has a count rate of $0.09\pm0.02$~counts~s$^{-1}$ (0.1-2.4~keV).

Second, TYC\;8241~2652~1 was in the field of view of a pointed \emph{ROSAT}/PSPC \citep{1987SPIE..733..519P} observation in 1993 with a total exposure time of 2.5 ks (Observation ID rp201497). It is detected with an excess of about 300 photons over the background (0.1-2.4~keV). This observation of TYC\;8241~2652~1 has not been analyzed before, and we present a spectral analysis in section~\ref{Xrayres}.

Third, TYC\;8241~2652~1 was observed three times in the \emph{XMM-Newton} slew survey \citep{2008A&A...480..611S} see table~\ref{tab:XMM}. \emph{XMM-Newton} (instrumental bandpass is 0.2-10~keV, but we restrict the analysis to 0.3 to 3~keV to reduce the background) continues to observe while the telescope slews from one target to the next. Data obtained serendipitously during these slews covers a large fraction of the sky, but with observations that often only last for a few seconds. Two slew exposures were taken before the IR flux dropped, and one of them during the IR flux decrease. \citet{2008A&A...480..611S} report that 90\% of all slew sources are found within 17\arcsec{} of their optical position, but larger offsets are possible. To ensure that all source photons are counted, even if they are not detected at the nominal position we extract counts in a circle with 30\arcsec{} radius centered on the optical position of TYC\;8241~2652~1, but we find that the results do not depend sensitively on the choice of extraction region. Background counts are determined from a region 100 times larger than the source extraction region. The background region is positioned in the slew path to have a comparable effective exposure time. Table~\ref{tab:XMM} shows the background counts normalized to the area of the source extraction region. Our target is seen with a total of 13 counts on a background of $< 1$ count; this is a significant detection.
Our target has a very soft spectrum in the slew data. Only 2 of 13 photons have energies $>1$~keV (1.1 and 2.2 keV). 
\begin{table*} 
\caption{{\emph{XMM-Newton} slew survey observations\label{tab:XMM}}}
    \begin{tabular}{ c c c c c c c}
        Slew Obs ID & Obs date & Exp. time & source & background & rate & flux\\ 
                    &          &  [s]      & [counts] & [counts / region] & [counts s$^{-1}$] & [$10^{-13}$ erg cm$^{-2}$ s$^{-1}$]\\
        \hline\hline
        9057600002 & 2003-02-31 & 10 & 6 & 0.53 & $0.6\pm0.2$ & $3\pm1$\\ 
        9111300003 & 2006-01-06 & 3 & 1 & 0.13 & $0.3\pm0.3$ & $2\pm2$\\ 
        9193600005 & 2010-07-06 & 7 & 6 & 0.13 & $0.9\pm0.3$ & $5\pm2$\\ 
        \hline
    \end{tabular} 
\end{table*}

Fourth, TYC 8241-2652-1 was observed with the \emph{Chandra} X-ray observatory on 2014-02-16 for 10~ks (ObsID 15713) 
with ACIS-S \citep{2003SPIE.4851...28G}, positioning the source on the back-illuminated chip S3 because of the better sensitivity at low energies (0.3-10~keV). To avoid pile-up from this relatively bright source, a subarray read-out mode was used. We processed all Chandra data with CIAO 4.8 \citep{Fruscione_2006} following the standard CIAO processing scripts for light curves and spectra. The background was determined from a large, source-free region on the same chip. Spectral fitting was done with the Sherpa fitting tool \citep{2007ASPC..376..543D}. The background in the observations is negligible for lightcurve and spectral extraction; pile-up does occur in TYC\;8241~2652~1, but it effects only 6\% of all detected events in the brightest pixel.

\subsection{Optical Spectroscopy}
\label{sect:optobs}
We continued monitoring activity signatures for TYC\,8241~2652~1 with measurements of H$\alpha$ in optical spectra. 

Two out of three new epochs of observations were performed with the WiFeS integral field unit \citep{2007Ap&SS.310..255D,2010Ap&SS.327..245D} on the Siding Spring Observatory 2.3\,m telescope. UT 03 April 2012 observations used the B3000 and R3000 low-resolution gratings (the number refers to the spectral resolving power for the given WiFeS grating) providing useable spectral data from 3400-8900\,\AA\ while UT 19 June 2013 observations used the B3000 low-resolution and R7000 high-resolution gratings providing complete coverage from 3400-7040\,\AA . WiFeS data were obtained in single-star mode with twice the spatial binning  (1$''$ spatial pixels). Spectra are obtained by optimally extracting and combining five image slices (effectively a 5$''$ diameter aperture around the object) that contain the majority of the stellar flux. Spectra are then relative flux-calibrated with observations of a flux standard star with known spectral response, except for the June 2013 red spectrum which is only continuum-normalized. Each spectrum has average signal-to-noise ratio per pixel $>$50.

The third epoch was obtained with the Goodman High Throughput Spectrograph on the 4.1m Southern Astrophysical Research (SOAR) Telescope at Cerro Pach\'{o}n. Observations were performed on UT 25 June 2014 and used the 0.46$''$ slit and the SYZY 400 lines\,mm$^{-1}$ grating resulting in spectral coverage from 4910-8960 \AA\ and a resolving power of $\approx$2000. The spectrum was continuum-normalized and has an average signal-to-noise ratio per pixel of $>$100.

\subsection{Near-infrared interferometry}
\label{sect:nearirobs}
Near-infrared interferometry provides an efficient way to reach the diffraction-limited resolution of single dish telescopes (with sparse aperture masking (SAM) interferometry) or to access even smaller spatial scales with long-baseline interferometry.  We observed TYC 8241 2652 1 both with SAM and long-baseline interferometry in order to search for close-in companions and to resolve any circumstellar material.

\subsubsection{VLTI/AMBER long-baseline interferometry}

In order to search for structures on astronomical unit scale around TYC 8241 2652 1, we employed VLTI/AMBER interferometry with the VLT 8.2m unit telescope triplet UT1-UT2-UT4. This telescope configuration allowed us to probe projected baseline lengths between 54 and 130\,m and position angles between 7 and $54^{\circ}$. The observations were conducted on 2012-12-24 as part of ESO observing programme 090.C-0697(A) (Kraus, Weigelt) and covered the near-infrared $K$-band ($2.20\pm 0.17\;\mu$m) with a spectral resolution of $\lambda / \Delta\lambda = 35$.
We recorded a total of 5000 interferograms on the target.  In order to calibrate the observables for atmospheric and instrumental effects, we bracketed the observations on TYC 8241 2652 1 with observations on the calibrator stars \object{HD42133} and \object{HD105316}, for which we adopt uniform disk diameters of $0.156 \pm 0.011$ and $0.109 \pm 0.008$\,mas, respectively (as computed with the JMMC SearchCal tool).  
To reduce the AMBER data, we employed the {\em amdlib} software \citep[Release 3;][]{tat07,che09} and extracted wavelength-dependent visibilities and phases as well as closure phases.

\subsubsection{VLT/NACO sparse aperture masking interferometry}

In order to reach the diffraction-limited resolution of the VLT 8.2\,m unit telescope UT4, we employed VLT/NACO Sparse Aperture Masking (SAM) interferometry.
The NACO data was recorded on 2012-12-19 as part of program 090.C-0904(A) (Cure, Kraus, Kanaan, Sitko, Ireland, Harries). We used NACO's 7-Hole mask and a near-infrared $L'$-band ($3.80 \pm 0.31\;\mu$m) filter.
During our two pointings on TYC 8241 2652 1, we recorded a total of 1110 interferograms with a detector integration time of 1\,s. The on-source observations were interlayed with observations on the calibrator star \object{HD105316} in order to calibrate instrumental closure phase effects.
The NACO data were reduced using our data reduction pipeline that was used already in various earlier studies \citep[e.g.][]{ire08,kra12,kra13}, providing absolute calibrated visibilities and closure phases.

\subsection{Mid-infrared imaging}
\label{sect:midirobs}
Ground-based mid-infrared imaging of TYC\,8241~2652~1 was pursued to continue monitoring the system and see if the disk material is beginning to return. Two epochs were obtained and are described below.

\subsubsection{Subaru/COMICS}

For a period between 2013-2015 inclusive, no mid-infrared imaging camera was reliably available in the 
Southern hemisphere and an attempt to 
detect TYC\,8241~2652~1 was made from Mauna Kea on UT June 19, 2013. 
Observations were performed with COMICS \citep{2000SPIE.4008.1144K,2003SPIE.4841..169O} on the Subaru 8.2\,m telescope in imaging mode with the 
N11.7 filter (11.7\,$\mu$m central wavelength and a width of 1.0\,$\mu$m). 
COMICS in imaging mode feeds a 320 $\times$ 240 pixel Si:As array whose 0.13$''$\,pixel$^{-1}$
plate scale affords a field of view of 41.6$''$ $\times$ 31.2$''$. Observations were chopped only (keeping all beams on chip
and lowering overhead) to maximize efficiency while obtaining sufficient sensitivity to detect faint sources. The chopping throw
used was 10$''$ along a position angle of 0$^{\circ}$ E of N. Data reduction follows standard high thermal background techniques. Chop pairs are differenced to remove the rapidly fluctuating background signal, then combined to yield the final reduced image. We
noted during the night that some sources displayed a slow vertical shift in one direction on the chip while being observed over
long periods of time; 
a variety of shift corrections were applied
to the TYC\,8241~2652~1 data during the reduction process with no impact on the final result. 
The flux calibration standard HD\,110458 \citep{1999AJ....117.1864C} was 
observed immediately prior to and immediately after observations of TYC\,8241~2652~1.

In practice, one cannot easily achieve the nominal instrumental sensitivity for COMICS when observing at an airmass of $\approx$3.
As such, it should not be surprising that TYC\,8241~2652~1 was not detected in the 800\,sec spent on source before it set below
the 15$^{\circ}$ elevation limit for Subaru. From the bracketed observations of HD\,110458 we estimate that the TYC\,8241~2652~1 
imaging sequence was capable of detecting sources as faint as 50\,mJy to $\approx$95\% confidence.

\subsubsection{VLT/VISIR}

As the Subaru observations were unable to provide the desired sensitivity to detect TYC\,8241~2652~1
at its low dust emission level, we waited for the VISIR instrument \citep{2004Msngr.117...12L,2015Msngr.159...15K}
at VLT-Melipal to become operational again before pursuing additional 
mid-infrared observations. Observations were performed in service mode on two consecutive nights in February, 2016 
in AutoChopNod imaging mode with default parameters, the chop/nod direction set to perpendicular, 
and blind positioning of the source in the upper left quadrant of the chip. 
The 1024 $\times$ 1024 pixel detector was configured for 
0.045\,$''$\,pixel$^{-1}$ yielding a field of view of roughly 46$''$ $\times$ 46$''$.
Observations were performed with the B10.7 filter (10.65\,$\mu$m central wavelength and a half-band width
of 1.37\,$\mu$m) and exposed for a total of 1525\,seconds on source. The flux standard 
HD 99167 \citep{1999AJ....117.1864C} was observed each photometric night immediately before observations of TYC\,8241~2652~1.
Observations on UT February 16, 2016 provided a weak detection of TYC\,8241~2652~1, but the seeing 
degraded after the start of the observation and did not meet the service mode specifications. As a 
result, observations were repeated on UT February 17, 2016 under better conditions.

Data reduction was performed within {\sf gasgano} using the standard {\sf VISIR\_img\_reduce} script.
For each separate night, images were chop- and nod-differenced and combined for both the standard star and 
TYC\,8241~2652~1. Flux was extracted for each positive and negative beam for both stars with an
aperture that yielded approximately 85\% encircled energy. Each of the four measurements were
averaged and the uncertainty set to the spread of these four measurements divided by 2 (the
square root of the number of measurements). When calculating the measured flux density of TYC\,8241~2652~1,
an additional 10\% uncertainty is added to the measurement of the flux calibration star's flux density to 
account for absolute calibration error and variable atmospheric conditions between observations of the two
sources. 
In practice this error is overshadowed by the relatively low signal-to-noise-ratio detections of 5-6 for for TYC\,8241~2652~1 which resulted in its individual detections having an uncertainty of about 20\%. We average the flux densities measured for each
night as they agree to within their 1$\sigma$ errors, yielding a final 10.7\,$\mu$m flux density of
11$\pm$2\,mJy.

\section{Results}
\label{sect:results}

\subsection{X-ray}\label{Xrayres}

In the \textit{Chandra} observation we detect two X-ray sources within 10\arcmin{} of aimpoint (at larger distances the point spread function becomes very large and the source identification becomes ambiguous). We identify the strongest source with TYC\;8241~2652~1. It is located $<0.5$\arcsec{} from the optical position, which is well within the 90\% error circle of \emph{Chandra}'s absolute pointing accuracy \footnote{\url{http://cxc.harvard.edu/proposer/POG/html/chap5.html}}; there are too few sources to improve the astrometry by comparison to other catalogs.

\begin{figure}[h!]
\begin{center}
\includegraphics[width=0.7\columnwidth]{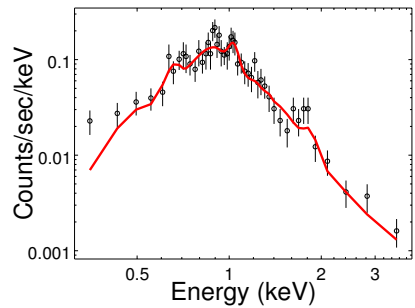}
\caption{{\emph{Chandra} spectrum binned to 20 counts per bin. The best-fit model is shown as red line. \label{fig:xray}%
}}
\end{center}
\end{figure}

The lightcurve of TYC\;8241~2652~1 is flat during the observation, indicating
that our analysis represents the quiescent state and not a stellar
flare. Stellar X-ray emission originates in collisionally ionized, optically
thin coronal plasmas, thus we fit a model of two thermal components
\citep[APEC,][]{2012ApJ...756..128F} and one absorption component for the
interstellar column density $n_\mathrm{H}$. A stellar accretion shock and
coronal plasma have very similar features in the spectrum and without grating
data we cannot measure the density predicted in shock models
\citep{2007A&A...466.1111G}; however the weak H$\alpha$ emission which just
fills the photospheric line (section~\ref{sect:opticalspectroscopy}) makes the
presence of an X-ray accretion shock very unlikely
anyway. Table~\ref{tab:xrayfit} shows the best-fit parameters. X-ray data and
the fit are presented in Fig.~\ref{fig:xray}. We use
\citet{1998SSRv...85..161G} as baseline for our abundances. The fit
  determines the ratio of observed to base abundance,
  e.g. $\frac{\textrm{Ne}_{\textrm{obs}}}{\textrm{Ne}_{\textrm{base}}}$.
However, the absolute abundances cannot be determined without grating
spectroscopy; if we e.g. multiplied all abundances in the model by
  three, the model would predict an almost identical spectrum. Thus, we
  normalize the abundances by fixing the abundance of elements with medium
first ionization potential (FIP) values (S, O, N, and C all have FIP between 10
and 15~eV) to 1. In other words, table~\ref{tab:xrayfit} gives e.g. \
$\frac{\frac{\textrm{Ne}_{\textrm{obs}}}{\textrm{Ne}_{\textrm{base}}}}{\frac{\textrm{O}_{\textrm{obs}}}{\textrm{O}_{\textrm{base}}}}$.

We combine Mg, Si, and Fe into one group, because they all have very similar FIP values (7.6-8.1~eV). The FIP of Ne is 21.6~eV.
The calculation of the emission measure (EM) and $L_\mathrm{X}$ uses $d = 120.9\pm5.4$~pc from \emph{GAIA data release 1} \citep{2016arXiv160904303L}; plasma temperatures are given as k$T$, where k is the Boltzman constant; and finally $L_\mathrm{X}$ is the intrinsic X-ray luminosity of the source in the 0.3-5.0~keV range.

\begin{table} 
    \caption{{Best-fit model for the X-ray spectrum ($1\sigma$ confidence intervals)\label{tab:xrayfit}}} 
    \begin{tabular}{ccc}
    \hline\hline
        parameter & soft comp. & hard comp.\\ 
        \hline
        k$T$ [keV] & $0.6\pm0.1$ & $1.7^{+0.7}_{-0.3}$ \\ 
        EM [$10^{52}$ cm$^{-3}$] & $6^{+5}_{-1}$ & $4^{+1}_{-3}$ \\ 
        Ne &  \multicolumn{2}{c}{$1.2^{+0.4}_{-0.6} $} \\ 
        Mg, Si, Fe &  \multicolumn{2}{c}{$ 0.2\pm0.1 $}\\ 
        $n_\mathrm{H}$ [$10^{20}$ cm$^{-2}$] &  \multicolumn{2}{c}{$ 0^{+3} $}\\ 
        \hline
        $L_\mathrm{X}$ [erg s$^{-1}$] & \multicolumn{2}{c}{$(1.1\pm0.3)\times10^{30}$ (unabsorbed, 0.3-5.0 keV)}\\
        \hline
    \end{tabular} 
\end{table}

The ROSAT pointed observation from 1993 presents a similar spectral result. We
find that TYC\;8241~2652~1 is detected with an excess of ca.\ 300 counts over
the background, corresponding to a count rate of 0.12 counts s$^{-1}$. As the
ROSAT PSPC detector has a much lower spectral resolution than Chandra, we
fitted the ROSAT spectrum with a coronal plasma model with a single temperature
component and solar elemental abundances \citep{1998SSRv...85..161G}. We find a
mean coronal temperature of k$T = 0.51$\,keV, i.e.\ somewhat lower than the
result from the Chandra observations. However this is not surprising, because
the ROSAT PSPC detector is only sensitive to photon energies from 0.1 to 2.4
keV, meaning that ROSAT generally sees the lower-temperature part of a stellar
corona. The X-ray luminosity found from the ROSAT spectral fit is similar to the Chandra observation: $L_\mathrm{X} = 1.4^{+0.1}_{-0.1}\times 10^{30}$\,erg/s extrapolated to an energy range of 0.3-5.0 keV to match the spectral band of the \textit{Chandra} observation.

We also calculate the X-ray luminosity for the ROSAT All-Sky Survey detection, using WebPIMMS to transform the detected count rate into an X-ray flux. We find an X-ray luminosity of $1.0^{+0.3}_{-0.3}\times 10^{30}$\,erg/s over an energy range of 0.3-5.0 keV.

For the \emph{XMM-Newton} slew survey data, the low count rate and the systematic errors from combining data from a moving source on three detectors cause large uncertainties on the measured count rate. Nevertheless, we attempt to estimate an energy flux from the count rate using WebPIMMS and the model fitted to \emph{ROSAT} data in Tab.~\ref{tab:XMM}.

\begin{figure}[h!]
\begin{center}
\includegraphics[width=0.7\columnwidth]{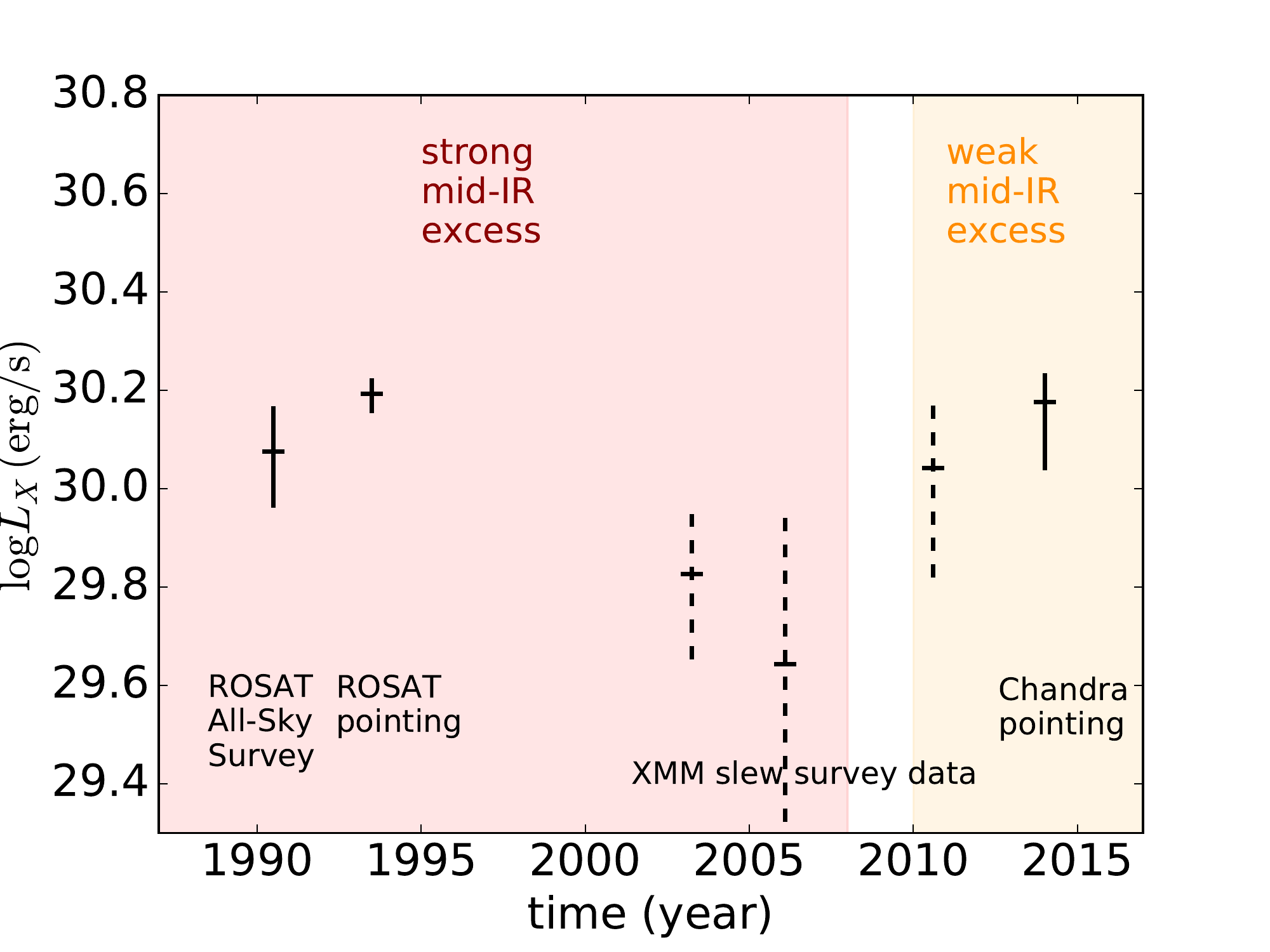}
\caption{{The observed X-ray luminosity of TYC\;8241~2652~1 is constant since 1991, with measurements before 1995 performed by ROSAT and the new measurement we report on here performed by Chandra in 2014. The low signal-to-noise detections achieved by the XMM-Newton slew survey from 2003 to 2010, indicated by dashed lines, are roughly consistent in luminosity with the previous and later X-ray observations.\label{LXlongterm}%
}}
\end{center}
\end{figure}

In summary, we find that the \textit{Chandra} spectrum and the total flux are fully compatible with the fluxes measured by \emph{ROSAT} and \emph{XMM-Newton}. There are no indications that the X-ray properties of TYC\;8241~2652~1 changed before and after it lost its disk; its X-ray luminosity is approximately constant over a timespan of $\sim$14 years, see Fig.~\ref{LXlongterm}.

Inactive stars with a weak corona commonly show an IFIP (inverse FIP) effect where elements with a low FIP like Fe, Si, and Mg have depleted abundances and elements with a high FIP (especially Ne) are enhanced \citep[see review by][and references therein]{2004A&ARv..12...71G}. The abundances measured in TYC\;8241~2652~1 in the \emph{Chandra} data follow exactly this pattern. Furthermore, active stars saturate around $\log L_\mathrm{X} / L_\mathrm{bol} = -3$ ($L_\mathrm{bol}$ is the bolometric luminosity); again, TYC\;8241~2652~1 matches this with $\log L_\mathrm{X} / L_\mathrm{bol} = -2.8$.

Compared to the X-ray temperatures of young stars on the one hand, e.g.\ T~Tauri stars observed in the Orion Ultradeep project (COUP) \citep{2005ApJS..160..401P}, TYC\;8241~2652~1 is at the lower end. The COUP sample contains actively accreting stars with circumstellar disks, as well as stars that have lost their disks already. Disk bearing stars often, but not always, have a stronger hot component \citep[e.g.][]{2010A&A...519A..97G}, compared with TYC\;8241~2652~1 where a similar amount of emission is seen in both components.

On the other hand, most debris disk systems are much older than T Tauri stars and so the typical range in $\log L_\mathrm{X} / L_\mathrm{bol}$ is -4.5 to -7 \citep{2013A&A...555A..11E}. 

\subsection{Optical Spectroscopy}

\label{sect:opticalspectroscopy}
H$\alpha$ equivalent widths are measured for the two WiFeS spectra reported herein.
We find that the H$\alpha$ line remains roughly critically filled in with an equivalent width
of 0$\pm$0.1 \AA\ in April 2012, consistent with the optical activity signature measurements reported by 
\citet{2012Natur.487...74M}. The June 2013 and June 2014 spectra show slightly enhanced optical activity in the H$\alpha$ line with equivalent widths of $-$0.8$\pm$0.1 \AA\ and $-$0.4$\pm$0.1 \AA{}, respectively (negative equivalent widths indicate emission above the continuum level).

With these optical spectra, and previous data and results reported by \citet{2012Natur.487...74M},
we estimate the visual extinction A$_V$ toward TYC\,8241~2652~1 to aid in interpreting X-ray
modeling results (see Sect.~\ref{sect:discussion}). 

From the 2MASS \citep{2006AJ....131.1163S} $K_s$-band measurement of $8.598\pm0.019$~mag, we get an
absolute K-band magnitude of $3.185\pm0.098$~mag. Note that this essentially
confirms the ~10 Myr old age of the source when compared against models from \citet{2002A&A...382..563B}.

The $V-K_s$ color of a star with spectral type K3V is 2.45 mags, thus we expect
$M_V=  5.635\pm0.098$~mag, and an expected apparent V-band magnitude
$m_v=11.048\pm0.138$~mag. This, when compared against the APASS \citep{2012JAVSO..40..430H} V-band magnitude of $m_v=11.118\pm0.035$ (taken when the dust was gone and hence appropriate for comparison to the Chandra measurement), suggests $A_V = 0.07\pm0.14$~mag.

\subsection{Near-infrared interferometry}

\begin{figure}[h!]
\begin{center}
\includegraphics[width=0.7\columnwidth]{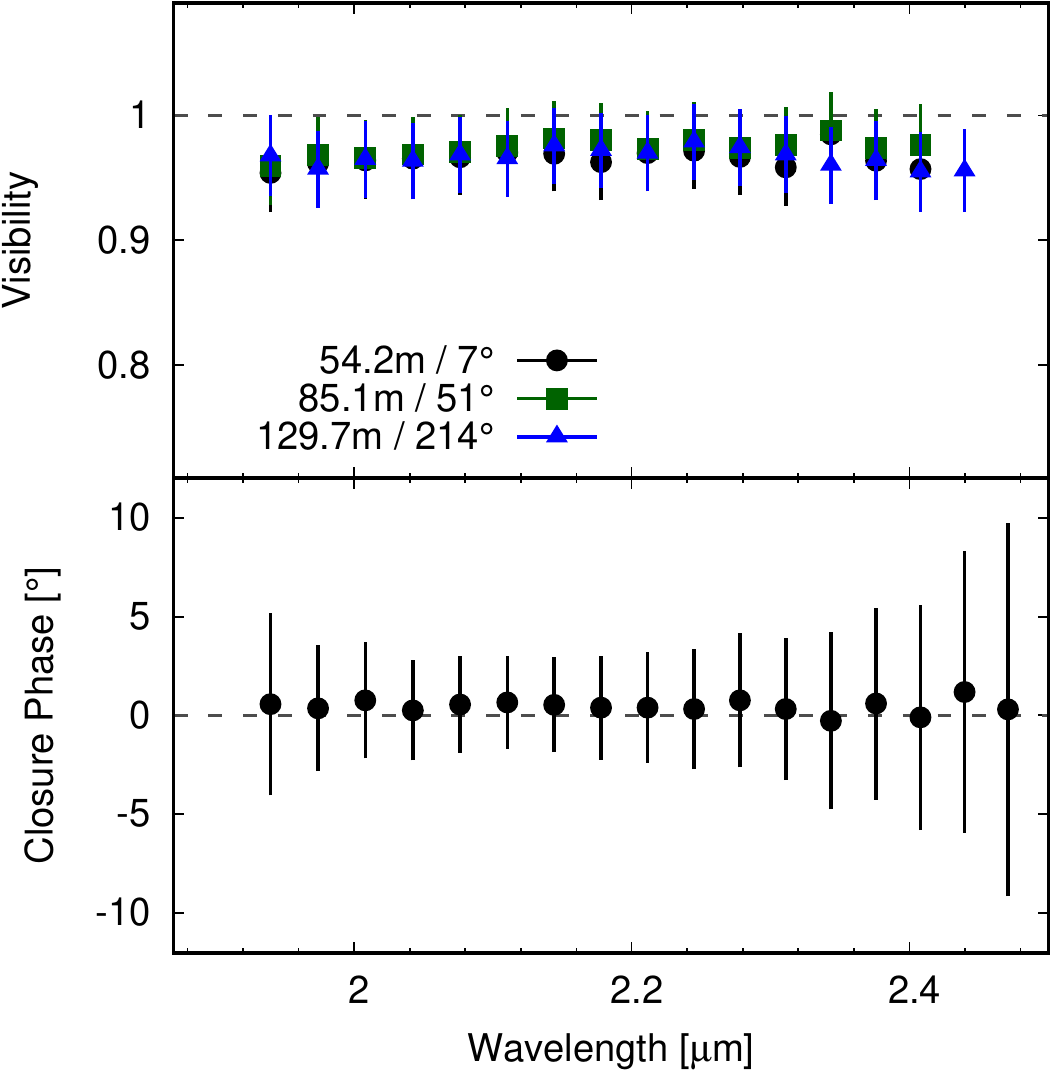}
\caption{{Visibilities (top) and closure phases (bottom) extracted from our VLTI/AMBER observations.\label{fig:AMBER}%
}}
\end{center}
\end{figure}

\begin{figure}[h!]
\begin{center}
\includegraphics[width=0.7\columnwidth]{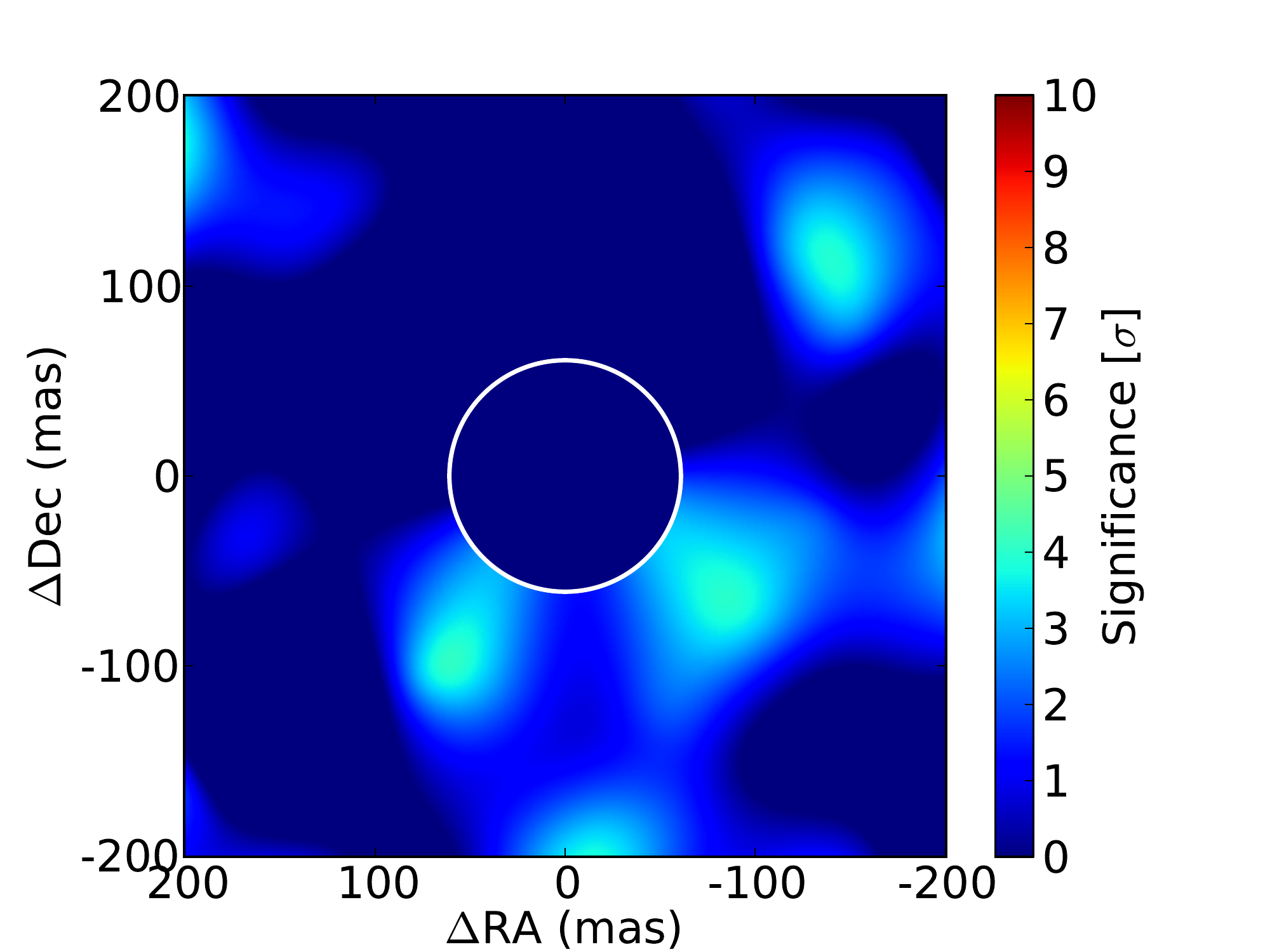}
\caption{{Significance map derived by fitting a companion model to our NACO $L'$-band SAM data. All structures in the map have a significance $\lesssim 4 \sigma$ in the 40-250\,mas range, which is consistent with a non-detection. The white inner circle represents our effective inner working angle with observations inside this annulus strongly adversely affected by SNR effects represented by limited contrast sensitivity. \label{fig:NACO}%
}}
\end{center}
\end{figure}

Our VLT/NACO SAM and VLTI/AMBER interferometric observations allow us to search for close-in companions in the separation range between $\sim 1$ and $250$\,milliarcseconds (mas), corresponding to physical scales of 0.12 ... 30\,AU at the distance of TYC 8241 2652 1.  

The AMBER observations probe the range from $\sim 1$ to 30\,mas in the $K$-band, while NACO covers the range from 20 to 250\,mas in the $L'$-band. Neither of the two data sets indicates any resolved emission. The measured visibilities are consistent with unity, indicating that there is no resolved $K$-band emission from circumstellar material on levels above $\sim 4$\% of the total flux (Fig.~\ref{fig:AMBER}).  

In order to search for asymmetries in the brightness distribution that might indicate the presence of a close-in companion, we modeled the measured closure phases with a companion model, where we fit the separation, position angle, and flux ratio between two point sources as free parameters. From the model fit we produced the significance map shown in Fig.~\ref{fig:NACO}. The map reveals some potential companion candidates in south-western direction at a separation of about 100\,mas -- however with a relatively low significance of $\lesssim 4 \sigma$. This significance is below our carefully established detection thresholds and we interpret the observation as a non-detection. We estimate that we would have been able to obtain clear detections for companions with contrasts lower than $\Delta L'=1.9$ between 20-40\,mas, $\Delta L'=3.6$ between 40-80\,mas, and $\Delta L'=4.4$ in the 80-240\,mas separation range.  The measured AMBER closure phases are also consistent with zero and we estimate that we can rule out companions $\Delta K=3$ in the 1-30\,mas range.

\subsection{Mid-infrared imaging}

\citet{2012Natur.487...74M} report depleted dust disk emission levels at $\approx$10\,$\mu$m for TYC\,8241~2652~1 of
12.6$\pm$0.3\,mJy in 2010 (from $WISE$) and 18$\pm$6\,mJy in 2012 (from Gemini/T-ReCS). These emission
levels are slightly elevated above the expected photospheric emission level of $\approx$11\,mJy
(see their Figure 1).
The COMICS non-detection and VISIR detection of TYC\,8241~2652~1 reveal that its circumstellar disk appears to have
maintained its depleted dust level since the 2010 $WISE$ measurements and has not yet returned to its pre-2009
exceptionally dusty state. Continued monitoring of the source is prudent as \citet{2012Natur.487...74M} suggest that the dust disk
might start to return within 10~years of its disappearance under some of their proposed models $-$ that anniversary is rapidly
approaching (sometime around 2019).

\section{Discussion}
\label{sect:discussion}
\subsection{Stellar activity and disk evaporation}
Cool stars generate magnetic fields with a dynamo analogous to the Sun. Our observations probe two manifestation of this stellar activity. The hot coronal gas is observed in the X-ray band, and the optical spectra probe the H$\alpha$ emission which originates in active regions in the stellar chromosphere.
Looking at the current X-ray spectrum as observed with \emph{Chandra} there are no indications that TYC\;8241~2652~1 differs in its X-ray properties from other young stars in any way. The $L_\textrm{X}/L_{\textrm{bol}}$ ratio and the H$\alpha$ equivalent width in TYC\;8241~2652~1 are very similar to what is seen on other young active stars such as T Tauri stars. For stars of this spectral type, an unperturbed photosphere shows H$\alpha$ in absorption, but in our observations, the photospheric H$\alpha$ line is filled, and shows at most weak emission above the continuum, indicating additional emission features from chromospheric activity. A H$\alpha$ equivalent width close to 0 is typical for active cool stars \citep{1990ApJS...74..891R} but much weaker than seen in accreting stars \citep{2003ApJ...592..282J}.
On the other hand, TYC\;8241~2652~1 is definitely more active than the average debris disk host star, consistent with the age of only 10~Myr \citep{2012Natur.487...74M}. 

The X-ray properties do not significantly differ before and after it lost its disk and the optical spectra before the mid-IR dimming \citep{2012Natur.487...74M} and after (this work) show consistent activity indicators.
Therefore, whatever happened to the disk apparently did not influence the stellar activity. However, stars as young as TYC\;8241~2652~1 typically spin so fast that they are in the saturated regime of stellar dynamo-powered magnetic activity. Therefore, even the accretion of large amounts of angular momentum and the accompanying spin-up of the star would not increase coronal and chromospheric activity and we cannot use activity to check if the stellar angular momentum increased through the quick accretion of the disk onto the star.

In six X-ray observations of TYC\;8241~2652~1 not a single flare is seen, but the summed observing time is still short. Thus, we look for a sample of comparable stars with a long enough observing time to estimate flare frequencies. By age and $L_X/L_\textrm{bol}$ value, TYC\;8241~2652~1 is comparable to T Tauri stars. The best sample, although still only about half as old as TYC\;8241~2652~1, is from the \emph{XMM-Newton} extended survey of the Taurus molecular cloud \citep[XEST;][]{2007A&A...468..353G}. Flares in this sample are analyzed by \citet{2007A&A...468..463S}. \citet{2013ApJ...765L..44O} estimate that a flare with a total energy of $10^{35}$~erg would be sufficient to explain the drop in IR luminosity in TYC\;8241~2652~1. While the flare itself would not do much to the dust, the coronal mass ejection that typically accompanies a flare should be powerful enough to remove a larger amount of dust from the system. In the XEST project, a flare of this magnitude was observed roughly every 800~ks (9 days) per star.

Individual coronal mass ejections on the Sun are confined to a small solid angle, so that not every large flare necessarily hits the disk; on the other hand, the coronal mass ejection in question would have to be big enough to influence a large fraction of the disk if invoked to explain the drop in IR luminosity between 2009 and 2010. If this mechanism worked even for a small fraction of the flares, we should have never observed an IR excess in TYC\;8241~2652~1, since its disk would be removed every few days. Obviously, there are large uncertainties in this argument, e.g. not all stars show the same level of activity, the XEST sample is younger than TYC\;8241~2652~1, and coronal mass ejection have not been directly detected for stars other than the sun \citep{2011A&A...536A..62L}.

\citet{2012Natur.487...74M} also discuss a scenario where an X-ray flare is responsible for the evaporation of dust in the disk. Their calculation is based only on the X-ray flux and they estimate a required flare energy about 10$^{39}$~erg - more than any stellar flare ever observed. The number $N$ of flares with energy larger than $E$ follows a power law of the form $\frac{dE}{dN} = k E^{-\alpha}$ \citep[review by][]{2004A&ARv..12...71G}. Taking the numbers reported in \citet{2007A&A...468..463S} for XEST, we would expect one flare with $E > 10^{39}$~erg per 100~Myr per star; obviously, we do not know if the extrapolation this far beyond any observed flare is valid, but it is certainly consistent with the fact that no other star has been observed to drop in the mid-IR luminosity like TYC\;8241~2652~1 \footnote{At face value, the probability to observe one such event when comparing the \emph{Spitzer} and \emph{WISE} catalogs, i.e. observing one decade in disk evolution for a few thousand young stars, is $10^{-5}$.}. \emph{Kepler} data from main-sequence stars indicates that the flare energy saturates around $2\times10^{37}$~erg \citep{2015ApJ...798...92W}. If such an upper limit exists for young stars, then no X-ray flare will ever be able to evaporate a disk like the one of TYC\;8241~2652~1 before 2009.

\subsection{Accretion and the dust/gas reservoir}

In principle, the X-ray $n_\mathrm{H}$, which probes the gas column density, can be compared with the optical reddening $A_V$, which probes the dust column density, to constrain the gas-to-dust ratio in the line-of-sight. However, the best fit is 0 for both values in TYC\;8241~2652~1.


\subsection{Companions and planets}
Our interferometry only found upper limits. Using the model tracks from \citet{2002A&A...382..563B,2003A&A...402..701B}, the flux limits can be converted to mass limits for any potential companion. The limit on $\Delta K$ in the 1-30 mas range (0.1-4~AU) implies $M < 0.2 M_{\odot}$; the upper limits on the mass of any companion derived from $\Delta L$ are 0.35, 0.1, and 0.05 $M_{\odot}$ in the region 20-40~mas (2-5~AU), 40-80~mas (5-10~AU), and 80-240~mas (10-30~AU), respectively. 
The collisional avalanche model and the runaway accretion model discussed in the supplementary material of \citet{2012Natur.487...74M} both require a sudden release of gas or dust. A collision of objects just 100-1000~km in size would be sufficient to start a collisional avalanche \citep{1997AREPS..25..175A,2007A&A...461..537G} and our observations do not pose limits for objects on this size scale. For completeness, we point out that the observed IR excess before 2009 might not have been a long-term stable disk, but could have been a remnant of the same collision that also provided the gas to start a runaway accretion event \citep[see the supplementary information of ][ for details and references]{2012Natur.487...74M}.

As an alternative, we want to describe ideas that would require a companion in
an eccentric orbit. This companion could have a cold disk itself, so cold that
it is only visible in the mid-IR when it passes TYC\;8241~2652~1 at a
distance $d$. At maximum luminosity the IR flux is 11\% of the flux of
TYC\;8241~2652~1 itself, so the disk of the companion needs to intercept at
least 11\% of the light from TYC\;8241~2652~1 and re-radiate it in the
IR. This translates to a circum-companion disk with radius $r_c >
\frac{d}{4.5}$. The temperature of the dust excess is not well determined
because it has not been detected above 30~$\mu$m.
\citet{2012Natur.487...74M} use 450~K as fiducal value, but
temperatures up to 800~K might be possible.
Given the flux from  TYC\;8241~2652~1, dust passing at 0.6~AU would be heated to 450~K.

\citet{2012Natur.487...74M} show 10$\mu$m fluxes from 2006, May 2008, and Jan 2009, peaking in 2008. The IR excess around TYC\;8241~2652~1 was also detected in 1983, suggesting an orbital period for the companion of 25 years, which implies a semi-major axis of $a=7.6$~AU.
In any case, the orbital motion is fastest at periastron, where the disk is heated most. Thus, orbital motion alone cannot explain why we see high emission for three years, but then a drop on a much faster time scale instead of a slow rise and fall. 
The faster time scale could be intrinsic to the companion system. If the disk precesses on a period of a few years, we can draw the following scenario: 
The disk was seen almost face-on around periastron in 2008. In 2010, the
companion was still close enough to have a hot disk, but the change in viewing
angle caused the drop in observable flux. In our new observations, the disk
might be face-on again, but is now, 8 years after periastron, sufficiently far
from the star that it is too cool to see. For example, an object with an orbit
with eccentricity $e=0.7$ spends about 4 years within 4~AU, but is beyond 11~AU 8 years after periastron.

An alternative to precessing the disk of the companion is to propose
that TYC\;8241~2652~1 itself has a cold, outer dust disk with an inner radius
of a few AU. This disk must be geometrically thin, such that it is not heated
significantly by the stellar radiation, but at the same time dense enough to be optically thick. Such a disk would be compatible
with the excess emission around 20~$\mu$~m detected by \emph{WISE} after the
main disk disappeared \citep{2012Natur.487...74M}. However, such a disk would likely also be dense enough to accrete onto the star, in contrast to our observations. The companion disk is heated near periastron and then vanishes from view as it is obscured by the larger disk of TYC\;8241~2652~1.

We can now calculate the minimum mass $m$ of the companion that is required to keep its disk gravitationally bound when passing TYC\;8241~2652~1 using the formula for the Hill radius from \citet{1992Icar...96...43H} and the distance of periastron passage $d= \frac{a (1-e^2)}{1+e}$:

\begin{equation}
\frac{m}{M_*} = 3\left(\frac{r_c}{a(1-e)}\right)^3 
= 3\left(\frac{\frac{d}{4.5}}{a(1-e)}\right)^3
=\frac{3}{4.5^3} = 0.03
\end{equation}

Note that this result is independent of orbital period and $a$. The minimum companion mass is just below our detection limit in the interferometry. Given that $r_c$ is a lower limit for a perfectly absorbing, face-on disk this scenario where TYC\;8241~2652~1 heats the disk of a passing secondary object seems unlikely, even if it is not formally ruled-out.

A more complex variant of the binary scenario requires only a cold, as yet unseen outer disk around TYC\;8241~2652~1, but no material around the companion. The companion induces turbulence in the disk and thus may contribute to heating it this way as it passes (our interferemetry limits are too strict to allow a companion with enough emission to heat the disk radiatively). Again, time scales are a problem. It seems unlikely that the turbulent heating can be sustained at an almost constant level for several years and then drops off suddenly.

\section{Summary}
\label{sect:summary}
In this article we present new X-ray observations, optical spectroscopy, near-IR interferometry, and mid-IR imaging of TYC\;8241~2652~1; all the data was taken to further explore physical scenarios that can explain how TYC\;8241~2652~1 reduced the dust mass in its disk by an order of magnitude or more between 2009 and 2010. The source is detected in new and archival X-ray observations. Lightcurves and spectra are fully consistent with typical values for young, active stars and there is no detectable change in X-ray activity before or after the dust mass loss in the disk. No absorbing column is detected towards the target. 
The interferometric data does not resolve any companion. In the inner region, it sets an upper limit of $0.2M_{\odot}$ on the mass of any potential binary component. Last, the mid-IR imaging detects a small excess above photospheric levels, which is consistent with the values observed just after TYC\;8241~2652~1 lost most of its dust in the disk.

This new data sets limits on several potential scenarios, but does not conclusively point to a single scenario for the dust loss.

\begin{acknowledgements}
We thank Michael Bessell from Mt. Stromlo/ANU and B.\ Bowler from Caltech for their help in obtaining and reducing optical spectra for this work. 
This research has made use of data obtained from the Chandra Data Archive, and software provided by the Chandra X-ray Center (CXC) in the application packages CIAO and Sherpa as well as data  obtained with XMM-Newton, an ESA science mission with instruments and contributions directly funded by ESA Member States and NASA. This research also used the AAVSO Photometric All-Sky Survey (APASS), funded by the Robert Martin Ayers Sciences Fund.
Support for H. M. G. was provided by the National Aeronautics and Space Administration through Chandra Award Number GO4-15009X issued by the Chandra X-ray Observatory Center, which is operated by the Smithsonian Astrophysical Observatory for and on behalf of the National Aeronautics Space Administration under contract NAS8-03060.
S.K.\ acknowledges support from an STFC Rutherford Fellowship (ST/J004030/1) and Marie Sklodowska-Curie CIG grant (Ref. 618910).
C.M.\ was supported by NASA grant 13-ADAP13-0178.
M.Cur{\'e} and S.Kanaan acknowledges financial support from Centro de Astrof\'isica de Valpara\'iso. S. Kanaan thank the support of Fondecyt iniciac\'ion grant N 11130702.
S.J.W. was supported by NASA contract NAS8-03060 (Chandra). We want to thank the referee for help specifically with the discussion section of the paper.
\end{acknowledgements}

\bibliography{bibliography/converted_to_latex.bib%
}

\end{document}